\documentclass[onecolumn]{aa}
\usepackage{graphicx}
\usepackage{txfonts}
\input{psfig.sty}
\begin{document}
   \title{Constraints on Chaplygin quartessence from the CLASS gravitational
lens statistics and supernova data}


   \author{Abha Dev
          \inst{1}
          \and
          Deepak Jain \inst{2}
          \and
          J. S. Alcaniz\inst{3}
          }

   \offprints{J. S. Alcaniz}

   \institute{Department of Physics and Astrophysics, University of Delhi,
Delhi110007, India\\
              \email{abha@ducos.ernet.in}
              \and
              Deen Dayal Upadhyaya College, University of Delhi, Delhi
110015, India\\
\email{deepak@physics.du.ac.in}
         \and
             Departamento de F\'{\i}sica, Universidade Federal do Rio
Grande do Norte, C.P. 1641, Natal - RN, 59072-970, Brasil\\
             \email{alcaniz@dfte.ufrn.br}
            }

   \date{Received ; accepted}

   \abstract{
The nature of the dark components (dark matter and dark energy)
that dominate the current cosmic evolution is a completely open
question at present. In reality, we do not even know if they
really constitute two separated substances. In this paper we use
the recent Cosmic All Sky Survey (CLASS) lensing sample to test
the predictions of one of the candidates for a unified dark
matter/energy scenario, the so-called generalized Chaplygin gas
(Cg) which is parametrized by an equation of state $p =
-A/\rho_{Cg}^{\alpha}$ where $A$ and $\alpha$ are arbitrary
constants. We show that, although the model is in good agreement
with this radio source gravitational lensing sample, the limits
obtained from CLASS statistics are only marginally compatible with
the ones obtained from other cosmological tests. We also
investigate the constraints on the free parameters of the model
from a joint analysis between CLASS and supernova data.

   }

   \maketitle
%

\section{Introduction}

As is well known, there is mounting observational evidence that
our Universe is presently dominated by two exotic forms of matter
or energy. Cold, nonbaryonic dark matter, which accounts for
$\simeq 30\%$ of the critical mass density and whose leading
particle candidates are the axions and the neutralinos, was
originally proposed to explain the general behavior of galactic
rotation curves that differ significantly from the one predicted
by Newtonian mechanics. Later on, it was also realized that the
same concept is necessary for explaining the evolution of the
observed structure in the Universe from density inhomogeneities of
the size detected by a number of Cosmic Microwave Background (CMB)
experiments. Dark energy or {\emph{quintessence}}, which accounts
for $\simeq 70\%$ of the critical mass density and whose leading
candidates are a cosmological constant $\Lambda$ and a relic
scalar field $\phi$, has been inferred from a combination of
astronomical observations which includes distance measurements of
type Ia supernovae (SNe Ia) indicating that the expansion of the
Universe is speeding up not slowing down (Perlmutter {\it et al.}
1999; Riess {\it et al.} 1998), CMB anisotropy data suggesting
$\Omega_{T} \simeq 1$ (de Bernardis {\it et al.} 2000; Spergel
{\it el al.} 2003), and clustering estimates providing $\Omega_m
\simeq 0.3$ (Calberg {\it et al.} 1996; Dekel, Burstein \& White
1997). While the combination of the last two results implies the
existence of a smooth component of energy that contributes with
$\simeq 2/3$ of the critical density, the SNe Ia results require
this component to have a negative pressure, which leads to a
repulsive gravity.

Despite the good observational evidence for the existence of these
two forms of energy, it has never been shown that in fact they
constitute two separate substances. In this concern, some authors
have proposed the so-called Unified Dark Matter/Energy scenarios
(UDME) or \emph{quartessence}, i.e., models in which these two
dark components are seen as different manifestations of a single
fluid (see, for instance, Matos \& Ure\~na-Lopez 2000; Davidson,
Karasik \& Lederer 2001; Watterich 2002; Kasuya 2001; Padmanabhan
\& Choudhury 2002). Among these theoretical proposals, an
interesting attempt of unification was originally suggested by
Kamenshchik {\it et al.} (2001) and developed by Bili\'c {\it et
al.} (2002) and Bento {\it et al.} (2002), namely, the Chaplygin
gas (Cg), an exotic fluid whose equation of state is given by
\begin{equation}
p_{Cg} = -A/\rho_{Cg}^{\alpha},
\end{equation}
with $\alpha = 1$ and $A$ a positive constant. In actual fact, the
above equation for $\alpha \neq 1$ constitutes a generalization of
the original Chaplygin gas equation of state proposed by Bento
{\it et al.} (2002). The idea of a dark-matter-energy unification
from an equation of state like Eq. (1) comes from the fact that
the Cg can naturally interpolate between nonrelativistic matter
($p = 0$) and negative-pressure ($p = -\rm{const.}$) dark energy
regimes (see Bento {\it et al.} 2002; Alcaniz, Jain \& Dev 2002
for details).

Very recently, there has been a wave of interest in exploring
theoretical (Bordemann \& Hoppe, 1993; Hoppe 1993; Jackiw 2000;
Gonzalez-Diaz 2003a; 2003b; Kremer 2003; Khalatnikov 2003; Balakin
{\it et al.} 2003; Bilic {\it et al.} 2003) and observational
consequences of the Chaplygin gas, not only as a possibility of
unification for dark matter/energy but also as a new candidate for
dark energy only. These models have been tested for a number of
cosmological data sets, including SNe Ia data (Fabris, Goncalves
\& de Souza, 2002; Colistete Jr. {\it et al.} 2003; Avelino {\it
et al.} 2003; Makler, de Oliveira \& Waga 2003), statistical
properties of gravitationally lensed quasars (Dev, Alcaniz \& Jain
2003; Silva \& Bertolami 2003), CMB measurements (Bento, Bertolami
\& Sen 2003a; 2003b; 2003c; Carturan \& Finelli 2002; Amendola
{\it et al.} 2003), age and angular size - redshift tests
(Alcaniz, Jain \& Dev 2002; Alcaniz \& Lima 2003), measurements of
X-ray luminosity of galaxy clusters (Cunha, Lima \& Alcaniz 2003),
future lensing and SNe Ia experiments (Avelino {\it et al.} 2003;
Silva \& Bertolami 2003; Sahni {\it et al.} 2003), as well as by
observations of large scale structure (Multamaki, Manera \&
Gaztanaga 2003; Bilic {\it et al.} 2003). The present situation is
somewhat controversial, with some tests indicating good agreement
between observational data and the theoretical predictions of the
model and others ruling out the model as an actual possibility of
description for our Universe (Sandvik {\it et al.} 2002; Bean \&
Dore 2003) (see, however, Be\c{c}a {\it et al.} 2003).

The aim of the current paper is to check the validity of such
models with radio-selected gravitational lens statistics and also
with a combination of gravitational lensing and SNe Ia data. To do
so, we adopt the most recent radio source gravitational lensing
sample, namely, the Cosmic All Sky Survey (CLASS) statistical data
which consists of 8958 radio sources out of which 13 sources are
multiply imaged (Browne {\it et al.} 2002; Chae {\it et al.}
2002). Here, however, we work only with those multiply imaged
sources whose image-splittings are known (or likely) to be caused
by single galaxies, which reduces the total number of lenses to 9.
For the cosmological background we consider a flat scenario in
which the generalized Cg together with the observed baryonic
content are responsible for the dynamics of the present-day
Universe (UDME or \emph{quartessence} models). In our computations
we adopt $\Omega_b = 0.04$, in accordance with the latest
measurements of the Hubble parameter (Freedman {\it et al.} 2002)
and of the baryon density at nucleosynthesis (Burles, Nollett \&
Turner 2001).

This paper is organized as follows. In Sect. 2 we present the
distance formulae necessary to our analysis. In Sect. 3 we discuss
the CLASS statistical sample, especially the observational
criteria used as well as the restrictions adopted in our analysis.
In Sect. 4, we derive the corresponding limits on Cg scenarios
from CLASS lensing statistics. We also examine the constraints
obtained from the statistical combination of lensing data with
recent SNe Ia observations and compare our constraints with others
derived from independent analyses. Finally, in Sect. 5, we finish
the paper by summarizing its main results.

\section{Basic equations}

By inserting Eq. (1) into the energy conservation law
($u_{\mu}T^{{\mu}{\nu}}_{;\nu} = 0$) one finds,
\begin{equation} \label{limB}
\rho_{Cg} = \left[A + B\left(\frac{R_o}{R}\right)^{3(1 +
\alpha)}\right]^{\frac{1}{1 +\alpha}},
\end{equation}
or, equivalently,
\begin{equation}
\rho_{Cg} = \rho_{Cg_{o}}\left[A_s + (1 -
A_s)\left(\frac{R_o}{R}\right)^{3(1+ \alpha)}\right]^{\frac{1}{1 +
\alpha}},
\end{equation}
where $\rho_{Cg}$ stands for the Cg energy density, the subscript
$o$ denotes present day quantities, $R(t)$ is the cosmological
scale factor, $B = \rho_{Cg_{o}}^{1 + \alpha} - A$ is a constant
and $A_s = A/\rho_{Cg_{o}}^{1 + \alpha}$ is a quantity related to
the sound speed of the Chaplygin gas today ($v_s^{2} = \alpha
A_s$).

A fundamental quantity related to the observables here considered is the
distance -redshift
relation, given by
\begin{equation}
\chi = \frac{c}{R_oH_o} \int_{(1 + z)^{-1}}^{1} {dx \over
x^{2} E(\Omega_{b}, A_s, \alpha, x)},
\end{equation}
where $x = {R(t) \over R_o} = (1 + z)^{-1}$ is a convenient
integration variable, $\Omega_{b}$ stands for the baryonic matter
density parameter, and the
dimensionless function $E(\Omega_{b}, A_s, \alpha, x)$ is written as
\begin{equation}
E = \left\{\frac{\Omega_{b}}{x^{3}} + (1 - \Omega_{b})\left[A_s + \frac{(1
-A_s)}{x^{3(\alpha + 1)}}\right]^{\frac{1}{\alpha + 1}}\right\}^{1/2}.
\end{equation}

For the lensing statistics developed in the next section, two
concepts are of fundamental importance, namely, the angular
diameter distance, $D_{LS}(z_L, z_S) = {R_or_1(z_L, z_S) \over (1
+ z_S)}$, between two objects, for example a lens at $z_L$ and a
source (galaxy) at $z_S$,
\begin{eqnarray}
D_{LS}(z_L, z_S)  =  \frac{c H_o^{-1}}{(1 + z_S)}
\times \int_{x'_S}^{x'_L} {dx \over x^{2} E(\Omega_{b}, A_s, \alpha, x)} .
\end{eqnarray}
and the age-redshift relation,
\begin{equation}
t_z = \frac{1}{H_o}\int^{(1 + z)^{-1}}_{0} {dx \over
x E(\Omega_{b}, A_s, \alpha, x)}.
\end{equation}

>From the above equations, we note that UDME models reduce to the
$\Lambda$CDM case for $\alpha = 0$ whereas  the standard Einstein-de Sitter
behavior is fully recovered for $A_s = 0$ (see also Fabris {\it et al.} 2003; Avelino
{\it et al.} 2003).

\section{Gravitational lensing statistics of the CLASS Sample}

Gravitational lensing directly probes the mass distribution in the
Universe so that an investigation of lensing events of sources at
high redshifts can provide important information about the global
cosmological parameters and the structure of the Universe. The use
of gravitational lensing statistics as a cosmological tool was
first considered in detail by Turner, Ostriker \& Gott (1984).
Subsequently, it was realized that a comparison of theoretical
lensing probabilities with gravitational lensing observations
provided an efficient constraint on the cosmological constant
(Fukugita, Futamase \& Kasai 1990; Turner 1990; Fukugita {\it et
al.} 1992; Kochanek 1996) or more generally on the density and the
equation of state of the dark energy component (Zhu 1998; 2000a;
2000b; Sarbu, Rusin \& Ma 2001; Chae {\it et al.} 2002; Huterer \&
Ma, 2003). However, the absence of an unbiased statistical sample
of sources that is complete to within well-defined observational
selection criteria and the uncertainties in the luminosity
function (LF) of galaxies have seriously complicated the
application of such methods.

\subsection{The CLASS Statistical Sample}

Recently, the CLASS collaboration \footnote{The Cosmic Lens All
Sky Survey (CLASS): http://www.aoc.nrao.edu/~smyers/class.html}
reported the so far largest lensing sample suitable for
statistical analysis, in which 13 out of the 8958 radio sources
are multiply imaged (Myers {\it et al.} 2002; Browne {\it et al.}
2002). This sample is well defined through the following
observational selection criteria (Myers {\it et al.} 2002; Chae
2002; Browne {\it et al.} 2002): (i) the spectral index between
1.4~GHz and 5~GHz is flatter than $-0.5$, i.e., \ $\alpha \ge
-0.5$ with $S_\nu\propto \nu^{\alpha}$, where $S_\nu$ is the flux
density measured in milli-jansky; (ii) the total flux density of
each source is $\ge$~20~mJy at 8.4~GHz; (iii) the total flux
density of each source is $\ge$~$S^0\equiv 30$~mJy at 5~GHz; (iv)
the image components in lens systems must have separations $\ge
0.3$~arcsec. The sources probed by CLASS at $\nu = 5$ GHz are well
represented by a power-law differential number-flux density
relation: $\left |dN/dS\right| \propto (S/S^{0})^{\eta}$ with
$\eta = 2.07 \pm 0.02$ ($1.97 \pm0.14$) for $S \geq S^{0}$ ($ \leq
S^{0}$). The redshift distribution of unlensed sources in the
sample is adequately described by a Gaussian model with a mean
redshift $z = 1.27$ and dispersion of $0.95$ (Chae 2002). Guided
by the above information about (i) the number-flux density
relation and (ii) the redshift distribution of unlensed sources,
we simulate the unlensed radio sources (8945 in number) of the
CLASS statistical sample using the Monte-Carlo technique
(rejection method).

In this paper, following Dev, Jain \& Mahajan (2003), we work only
with those multiply imaged sources whose image-splittings are
known (or likely) to be caused by single galaxies. This means that
our database is constituted by 9 lenses out of a sample of 8954
radio sources.

\subsection{Lensing Statistics}

We start our analysis by assuming the singular isothermal sphere (SIS)
model for the lens mass distribution. As has been discussed elsewhere this
assumption is a good approximation to the real mass distribution in galaxies
(see, e.g., Turner, Ostriker \& Gott 1984). In this case, the cross-section for lensing
events
is given by\begin{equation}
\sigma_{\rm SIS} = 16\pi^3 (\frac{v}{c})^4(\frac{D_{OL} D_{LS}}{D_{OS}})^2,
\end{equation}
where $v$ represents the velocity dispersion and $D_{OL}$,
$D_{OS}$ and $D_{LS}$ are, respectively, the angular diameter
distances from the observer to the lens, from the observer to the
source and between the lens and the source.By ignoring evolution
of the number density of galaxies and assuming that the comoving
number density is conserved, the differential probability of a
lensing event can be expressed as
\begin{equation}
d\tau = n_o(1 + z_L)^{3}\sigma_{\rm{SIS}}\frac{cdt}{dz_L}dz_L ,
\end{equation}
where the quantity $dt/dz_L$ can be easily obtained from Eq. (7)
and the present-day comoving number density of galaxies is
\begin{equation} \label{number}
n_o = \int_0^{\infty} \phi(L) dL.
\end{equation}
The differential optical depth of lensing in traversing $dz_L$ with angular
separation between $\phi$ and $\phi + d\phi$ is (Fukugita, Futamase \& Kasai 1990;
Turner 1990; Fukugita {\it et al.} 1992):
\begin{eqnarray}
\frac{d^{2}\tau}{dz_{L}d\phi}d\phi dz_{L}
= F^{*}\,(1 + z_{L})^{3}\,\left({{D_{OL} D_{LS}}\over{ R_{o}
D_{OS}}}\right)^{2}\,\frac{1}{R_{o}}\, \frac{cdt}{dz_{L}} \times \,
\frac{\gamma/2}{\Gamma(\alpha+1+\frac{4}{\gamma})}
\left(\frac{D_{OS}}{D_{LS}}\phi\right)^{\frac{\gamma}{2}(\alpha+1+\frac{4}{
\gamma})} \times
\,{\rm{exp}}\left[-\left(\frac{D_{OS}}{D_{LS}}\phi\right)^{\frac{\gamma}{2}
}\right]\frac{d\phi}{\phi}
dz_{L},
\label{diff}
\end{eqnarray}
where the function $F^{*}$ is defined as
\begin{equation}
F^* = {16\pi^{3}\over{c\, H_{0}^{3}}}\phi_\ast v_\ast^{4}\Gamma\left(\alpha
+{4\over\gamma}
+1\right).
\end{equation}
In Eq. (\ref{number}), $\phi (L)$ is the Schechter LF (Schechter
1976) given by
\begin{equation}
\phi(L)\,dL=\phi_*\left (L\over L_*\right )^\alpha
\exp(-L/L_*) {dL\over L_*}.
\label{eq:LF}
\end{equation}
In order to relate the characteristic luminosity $L_*$ to the
characteristic velocity dispersion $v_{*}$, we use the Faber-Jackson
relation (Faber \& Jackson 1976) for E/S0 galaxies ($L_* \propto {v_*}^{\gamma}$),
with $\gamma = 4$. For the analysis presented here we neglect the
contribution of spirals as lenses because their velocity dispersion is small
when compared to ellipticals.

The two large-scale galaxy surveys, namely, the 2dFGRS
\footnote{The 2dF Galaxy RedshiftSurvey (2dfGRS):
http://msowww.anu.edu.au/2dFGRS/} and the SDSS \footnote{ Sloan
Digital Sky Survey: http://www.sdss.org/} have produced converging
results on the total LF. The surveys determined the Schechter
parameters for galaxies (all types) at $z \le 0.2$. Chae (Chae
2002) has worked extensively on the information provided by these
recent galaxy surveys to extract the local type-specific LFs. For
our analysis here, we adopt the normalization-corrected Schechter
parameters of the 2dFGRS survey (Folkes {\it et al.} 1999):
$\alpha = -0.74$, $\phi^{*} = 0.82 \times 10^{-2} h^3
\mathrm{Mpc^{-3}}$, $v^{*} = 185\,\mathrm{km/s}$ and $F^{*} =
0.014$.

The normalized image angular separation distribution for  a source at
 $z_{S}$ is obtained by integrating $\frac{d^{2}\tau}{dz_{L}\,d\phi}$
over $z_{L}$:

\begin{equation}
{d{\mathcal{P}}\over d\phi}\, =\,
{1\over\tau(z_S)}\int_{0}^{z_{s}}\,{\frac{d^{2} \tau }{dz_{L}d\phi}}
{dz_{L}}.
\end{equation}

The corrected (for magnification and selection effects) image separation
distribution function for a single source at redshift $z_{S}$ is given by
(Kochanek 1996; Chiba \& Yoshii 1999)
 \begin{eqnarray}
 P'(\Delta\theta)\, = \, \mathcal{B}\,{{{\gamma}} \over
 2\, \Delta \theta} \int_{0}^{z_S}\left [{D_{0S}\over{D_{LS}}} \phi
 \right ]^{{\frac{\gamma}{2}}(\alpha + 1+ {\frac{4}{\gamma}})}
 \,F^{*}\,{cdt\over dz_{L}} \times \exp\left[- \,\left({D_{0S}\over{D_{LS}}}
\phi\right)^{\frac{\gamma}{2}}\right
 ] {(1 + z_{L})^{3} \over
\Gamma\left(\alpha +{4\over\gamma} +1\right)} \times  \left[\,\left ({D_{OL}D_{LS}\over
R_0
  D_{OS}}\right)^{2}\,{1\over R_0}\right ]\,\,{dz_{L}}.
\label{dist}
 \end{eqnarray}

\noindent Similarly, the corrected lensing probability for a given
source at redshift $z$ is given  by
\begin{equation}
P' = \tau(z_S) \,\int {d{\mathcal{P}}\over d\phi} \mathcal{B}\; d\phi.
\label{prob}
\end{equation}
Here $\phi$ and $\Delta\theta$ are related as $\phi =
{\frac{\Delta\theta}{8 \pi (v^{*}/c)^2}}$, and $\mathcal{B}$ is
the magnification bias. This is taken into account because, as
widely known, gravitational lensing causes a magnification of
images and this transfers the  lensed sources to higher flux
density bins. In other words, the lensed sources are
over-represented in a flux-limited sample. The magnification bias
${\mathcal B}(z_S,S_\nu)$ increases the lensing probability
significantly in a bin of total flux density ($S_\nu$) by a factor
\begin{eqnarray}
\mathcal{B}(z_S,S_\nu) =  \left |\frac{dN_{z_S}(>S_\nu)}{dS_\nu}
\right|^{-1} \times \int_{\mu_{min}}^{\mu_{max}} \left
|\frac{dN_{z_S}(>S_\nu/\mu)}{dS_\nu}\,p(\mu)\right |{1\over\mu}
\,d\mu.
\label{B2}
 \end{eqnarray}
Here  $N_{z_{S}}(> S_\nu)$ is the intrinsic flux density relation
for the source population at redshift $z_{S}$. $N_{z_{S}}(>
S_\nu)$ gives the number of sources at redshift $z_{S}$ having
flux greater than $S_\nu$. For the SIS model, the magnification
probability distribution is $p(\mu) = 8/{\mu}^{3}$. The minimum
and maximum total magnifications $\mu_{min}$ and $\mu_{max}$ in
Eq. (\ref{B2}) depend on the observational characteristics as well
as on the lens model. For the SIS model, the minimum total
magnification is $\mu_{min} \simeq 2$ and the maximum total
magnification is $\mu_{max} = \infty$. The magnification bias
$\mathcal B$ depends on the differential number-flux density
relation $\left|dN_{z_{S}}(> S_{\nu})/dS_{\nu}\right|$. The
differential number-flux relation needs to be known as a function
of the source redshift. At present redshifts of only a few CLASS
sources are known. We, therefore, ignore redshift dependence of
the differential number-flux density relation. Following Chae
(2002), we further ignore the dependence of the differential
number-flux density relation on the spectral index of the source.

\begin{figure}
\vspace{.10in}
\centerline{\psfig{figure=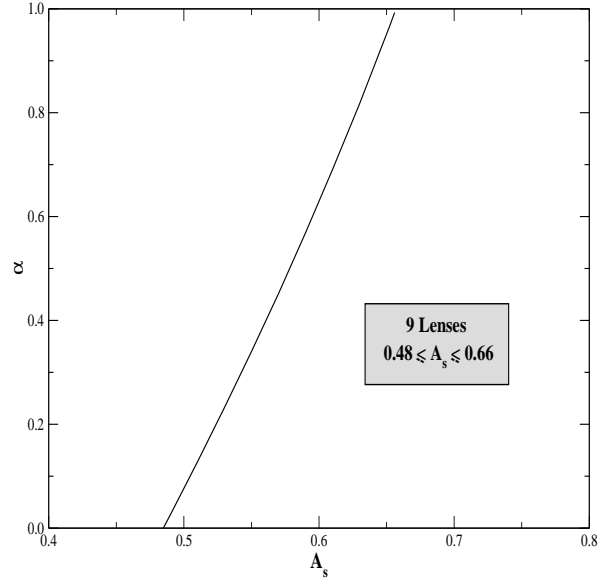,width=3.2truein,height=3.2truein
,angle=-90}\hskip 0.1in}
\caption{Contour for 9 lensed radio sources in the parametric space $A_s -
\alpha$. While the entire range of $\alpha$ is allowed, the parameter $A_s$
is restricted to the interval $0.48 \leq A_s \leq 0.66$.}
\end{figure}

An important selection criterion for the CLASS statistical sample
is that the ratio of the flux densities of the fainter to the
brighter images ${\mathcal R}_{min}$ is $ \ge 0.1$. Given such an
observational limit,the minimum total magnification for double
imaging for the adopted model of the lens is (Chae 2002):
\begin{equation}
\mu_{min} = 2 \frac{1+\mathcal{R}_{min}}{1-\mathcal{R}_{min}}.
\end{equation}
Another selection criterion is that the image components in lens
systems must have separations $\ge 0.3$~arcsec. We incorporate
this selection criterion by setting the lower limit of $\Delta
\theta$ in Eq. (\ref{prob}) as $0.3$ arcsec.

\section{Testing Cg Scenarios against Observations}

The expected number of lensed radio sources is $N_{\rm lens} =
\sum P'_{i}$, where $P'_{i}$ is the lensing probability of the
$i^{th}$ source and the sum is over the entire adopted sample. The
expected number of lensed sources is thus a function of the
parameters $A_s$ and $\alpha$. We have done a grid search for
those combinations ($A_s$, $\alpha$) by fixing $N_{\rm lens} = 9$.
In Fig. 1 we show the contour for 9 lensed radio sources in the
parametric space $A_s - \alpha$. As can be seen, while the entire
range of $\alpha$ is allowed, the parameter $A_s$ is tightly
restricted to the interval $0.48 \leq A_s \leq 0.66$. This
particular range for $A_s$ is not compatible with the one obtained
from a SNe analysis involving 92 events of the Supernova Cosmology
Project and High-$z$ supernova Search Team, i.e., $A_s > 0.69$ at
95\% confidence level (Avelino {\it et al.} 2003) and is only
marginally compatible with the limits from age + SNe performed by
Makler, de Oliveira \& Waga (2003). A comparison between the above
interval with the one restricted by age estimates of high-$z$
objects shows that the $N_{\rm lens}$ test for the CLASS sample is
compatible with the existence of the radio galaxy LBDS 53W091 (3.5
Gyr at $z = 1.55$) which implies $A_s \geq 0.52$ but that it is
not in accordance with the existence of the 4.0-Gyr-old radio
galaxy 53W069 (at $z =1.43$) and the 2.0-Gyr-old quasar APM
08279+5255 (at $z = 3.91$) which requires, respectively, $A_s \geq
0.72$ and $A_s \geq 0.82$ (Alcaniz, Dev \& Jain 2003). The above
interval from the $N_{\rm lens}$ test is also not in agreement
with the tight limits obtained by Silva \& Bertolami (2003) from
future SNe and lensing data, i.e., $0.75 \leq A_s \leq 0.79$ at
2$\sigma$ ($\alpha \lesssim 0.2$).

The likelihood function for lensing can be written as
\begin{equation}
{\cal{L}} = \prod_{i=1}^{N_{U}}(1-P^{'}_{i})\,\prod_{k=1}^{N_{L}}
P'(\Delta\theta).
\label{LLF}
\end{equation}
Here $N_{L}$ is the {\it observed} number of multiple-imaged
lensed radio sources and $N_{U}$ is the number of unlensed sources
in the adopted sample. The results of our analysis for the Cg
model are displayed in Fig. 2a. The contours correspond to the
$68.3\%$ and $95.4\%$ confidence level (cl) in the
($A_s$,$\alpha$) plane. Although the entire range of $\alpha$ is
permissible within the 68.3\% confidence level, the parameter
$A_s$ is constrained to be $\leq 0.8$ at $68.3\%$ cl and $\leq
0.9$ at $95.4 \%$ cl. For this analysis the best fit model occurs
for $A_s=0.36$ and $\alpha=0.35$, which corresponds to an
accelerating scenario with a deceleration parameter $q_o = -0.02$
and a total expanding age of $7.33h^{-1}$ Gyr. Although not very
restrictive, the constraints on the parameter $A_s$ from the CLASS
lensing sample of radio sources are more stringent than those
obtained from the optical gravitational lensing surveys for
quasars (Dev {\it et al.} 2003). However, it still limits the
parameter $\alpha$ very weakly.
\begin{figure}
\vspace{.10in}
\centerline{\psfig{figure=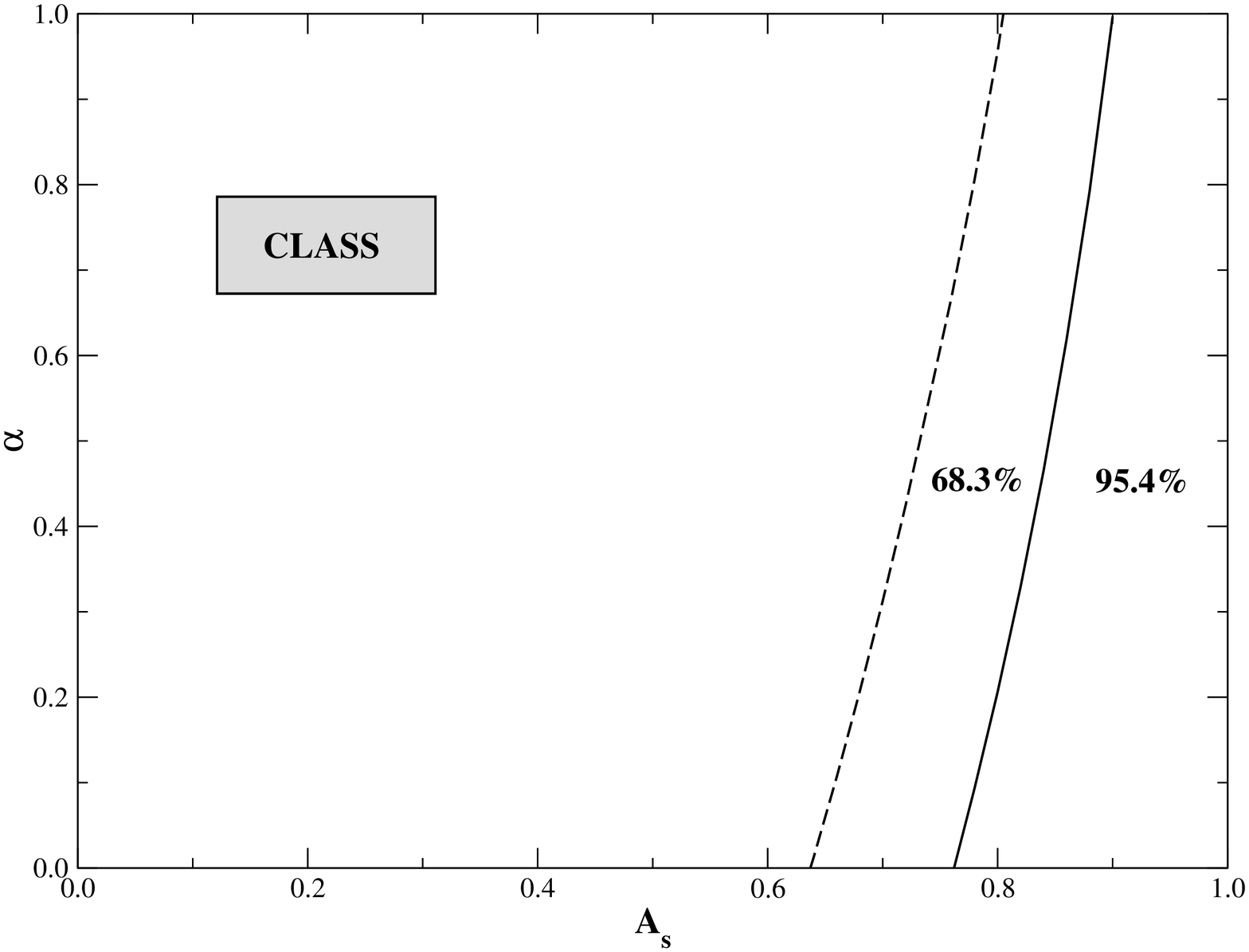,width=3.2truein,height=3.2truein,angle=-90}
\psfig{figure=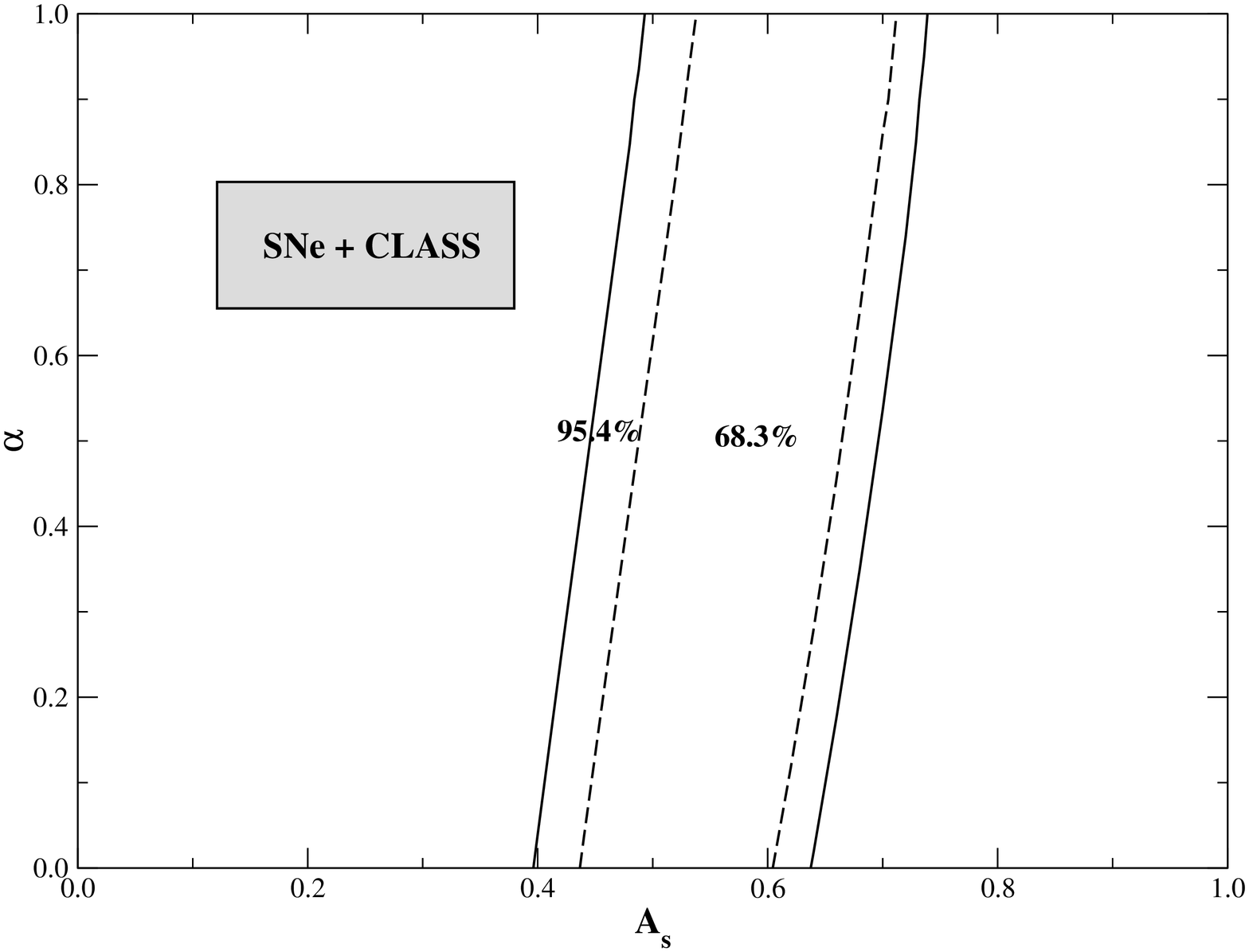,width=3.2truein,height=3.2truein,angle=-90}
\hskip 0.1in}
\caption{{\bf{a)}} Confidence regions in the plane $A_s - \alpha$ arising from
CLASS lensing statistics. Dashed and solid lines indicate contours of
constant likelihood at $68.3\%$ and $95.4\%$ confidence levels, respectively. {\bf{b)}}
The likelihood contours in the $\alpha - A_s$ plane for the joint
lensing + SNe Ia analysis described in the text.}
\end{figure}

\subsection{Joint analysis with supernova data}

In order to perform a joint analysis with CLASS and SNe data sets,
we first follow the conventional magnitude-redshift test (see, for
example, Goliath {\it et al.} 2001; Dicus \& Repko 2003;
Padmanabhan \& Choudhury 2003) and use the SNe Ia data set that
corresponds to the primary fit C of Perlmutter {\it et al.} (1999)
together with the highest redshift supernova observed so far, i.e,
the 1997ff at $z = 1.755$ and effective magnitude $m^{eff} =26.02
\pm 0.34$ (Benitez {\it et al.} 2002) and two newly discovered SNe
Ia, namely, SN 2002dc at $z= 0.475$ and $m^{eff} =22.73 \pm 0.23$
and SN 2002dd at $z = 0.95$ and $m^{eff} = 24.68\pm 0.2$
(Blakeslee {\it et al.} 2003). We thus work with a total of $57$
supernovae. The apparent magnitude of a given SNe is related to
the luminosity distance $d_L$ by the known relation $m =
\mathcal{M} + 5\log D_L$, where $D_L = H_0\,d_L $ and
$\mathcal{M}$ is given by the intercept obtained by fitting the
low-redshift data set to $m(z) = \mathcal{M} + 5\log (cz)$ (Hamuy
{\it et al.} 1996). The value obtained is ${\mathcal{M}} = -3.325$
and confirms the results of Perlmutter {\it et al.} (1999). For
the sake of completeness, we perform such SNe analysis. The best
fit model occurs for $A_s = 0.52$ and $\alpha = -0.2$ with a
minimum value of $\chi^2_{\mathrm {min}} = 64.31$ (which
corresponds to $\chi^2 _\nu = 1.17$). When this magnitude-redshift
test is combined with CLASS lensing statistics, tighter
constraints on the $A_s$ parameter can be obtained. As was shown,
the index $\alpha$ is highly insensitive to SNe Ia data (Makler,
de Oliveira \& Waga 2003). Figure 2b shows the result of this
joint analysis. For the combined $\chi^{2}$ analysis we used
$\chi^{2}_{total} = \chi^{2}_{SNe} - 2{\rm{ln}}l$, where $l =
{\cal{L}}_{lens}/{\cal{L}}_{max}^{lens}$ is the normalized
likelihood for lenses. As can be seen, the limits on $A_s$ are
more restrictive now than those imposed by the gravitational
lensing statistics of the CLASS sample (Fig. 2a). Within 68.3\%
cl, the constraints on the parameter $A_s$ are as follows: $0.39
\leq A_s \leq 0.71$ at 68.3\% cl and  $0.35 \leq A_s \leq 0.74$ at
95.4\% cl. In particular, the best fit model occurs for $A_s =
0.58$ and $\alpha = 0.5$, corresponding to a $7.93h^{-1}$-Gyr-old,
accelerating Universe with a deceleration parameter $q_o = -0.33$.

\section{conclusion}

A considerable amount of observational evidence suggests that the
current evolution of our Universe is fully dominated by two dark
components, the so-called dark matter and dark energy. The nature
of these components, however, is a tantalizing mystery at present,
and it is not even known if they constitute two separate
substances. In this paper we have investigate some observational
predictions of cosmologies driven by an exotic component named the
generalized Chaplygin gas. These models constitute an interesting
possibility of unification for dark matter/energy, where these two
dark components are seen as different manifestations of a single
fluid (UDME). We have investigated observational constraints from
lensing statistics on spatially flat UDME scenarios. Since
gravitational lensing statistics constitutes an independent way of
constraining cosmological parameters we have used the most recent
lensing data, namely, the Cosmic All Sky Survey (CLASS) sample to
obtain the $68.3\%$ and $95.4\%$ confidence intervals on the
parameters of the Cg equation of state. Our statistical analysis
shows that the best fit scenario for these data occurs at
$A_s=0.36$ and $\alpha=0.35$. At $68.3\%$ cl, parameter $A_s$ is
restricted to $\leq 0.8$ while the entire range of $\alpha$ is
allowed. By considering the observed number of lensed radio
galaxies we tightly constrain $A_s$ to the interval $0.48 \leq A_s
\leq 0.66$. From a joint $\chi^{2}$ analysis with SNe Ia data we
obtain $0.35 \leq A_s \leq 0.74$ at 95.4\% cl with the best fit
model occurring for $A_s = 0.58$ and $\alpha = 0.5$, which
corresponds to an accelerating scenario with a deceleration
parameter $q_o = -0.33$ and a total expanding age of 7.93$h^{-1}$
Gyr. As has been commented earlier, such results are only
marginally consistent with those obtained from independent
cosmological tests. It means that only with a more general
analysis, possibly a joint investigation involving different
classes of cosmological data, it will be possible to delimit the
$A_s - \alpha$ plane more precisely, as well as to test more
properly the consistency of these scenarios as a viable
possibility of unification for the dark matter and dark energy
scenarios.


\begin{acknowledgements}
The authors are very grateful to Zong-Hong Zhu and R. Silva for many valuable
discussions and a critical reading of the manuscript. JSA is supported by
the Conselho Nacional de Desenvolvimento Cient\'{\i}fico e Tecnol\'{o}gico
(CNPq - Brasil) and CNPq(62.0053/01-1-PADCT III/Milenio).
\end{acknowledgements}

\end{document}